\documentclass[journal]{IEEEtran}

\ifCLASSINFOpdf
\else
\fi

\usepackage{graphicx,graphics,epsfig,epstopdf,float}
\usepackage{amsmath} 
\usepackage{amssymb}  
\usepackage{mathrsfs}
\usepackage{enumerate}
\usepackage{subfigure}
\usepackage{booktabs}
\usepackage{colortbl}
\usepackage{color}  
\usepackage{bm}
\usepackage{psfrag}
\usepackage{cite}
\usepackage{algorithm}
\usepackage{algorithmicx}
\usepackage{algpseudocode}
\usepackage{mdwlist}
\usepackage{longtable}
\usepackage{array}
\usepackage{acronym}  
\usepackage{svg}
\usepackage{bm}

\DeclareMathAlphabet{\mathsfbr}{OT1}{cmss}{m}{n}
\SetMathAlphabet{\mathsfbr}{bold}{OT1}{cmss}{bx}{n}
\DeclareRobustCommand{\msf}[1]{%
  \ifcat\noexpand#1\relax\msfgreek{#1}\else\mathsfbr{#1}\fi
}

\makeatletter
\newcommand{\msfgreek}[1]{\csname s\expandafter\@gobble\string#1\endcsname}
\makeatother

\DeclareFontEncoding{LGR}{}{} 
\DeclareSymbolFont{sfgreek}{LGR}{cmss}{m}{n}
\SetSymbolFont{sfgreek}{bold}{LGR}{cmss}{bx}{n}
\DeclareMathSymbol{\salpha}{\mathord}{sfgreek}{`a}
\DeclareMathSymbol{\sbeta}{\mathord}{sfgreek}{`b}
\DeclareMathSymbol{\sgamma}{\mathord}{sfgreek}{`g}
\DeclareMathSymbol{\sdelta}{\mathord}{sfgreek}{`d}
\DeclareMathSymbol{\sepsilon}{\mathord}{sfgreek}{`e}
\DeclareMathSymbol{\szeta}{\mathord}{sfgreek}{`z}
\DeclareMathSymbol{\seta}{\mathord}{sfgreek}{`h}
\DeclareMathSymbol{\stheta}{\mathord}{sfgreek}{`j}
\DeclareMathSymbol{\siota}{\mathord}{sfgreek}{`i}
\DeclareMathSymbol{\skappa}{\mathord}{sfgreek}{`k}
\DeclareMathSymbol{\slambda}{\mathord}{sfgreek}{`l}
\DeclareMathSymbol{\smu}{\mathord}{sfgreek}{`m}
\DeclareMathSymbol{\snu}{\mathord}{sfgreek}{`n}
\DeclareMathSymbol{\sxi}{\mathord}{sfgreek}{`x}
\DeclareMathSymbol{\somicron}{\mathord}{sfgreek}{`o}
\DeclareMathSymbol{\spi}{\mathord}{sfgreek}{`p}
\DeclareMathSymbol{\srho}{\mathord}{sfgreek}{`r}
\DeclareMathSymbol{\ssigma}{\mathord}{sfgreek}{`s}
\DeclareMathSymbol{\stau}{\mathord}{sfgreek}{`t}
\DeclareMathSymbol{\supsilon}{\mathord}{sfgreek}{`u}
\DeclareMathSymbol{\sphi}{\mathord}{sfgreek}{`f}
\DeclareMathSymbol{\schi}{\mathord}{sfgreek}{`q}
\DeclareMathSymbol{\spsi}{\mathord}{sfgreek}{`y}
\DeclareMathSymbol{\somega}{\mathord}{sfgreek}{`w}

\DeclareMathSymbol{\svarsigma}{\mathord}{sfgreek}{`c}

\DeclareMathSymbol{\sGamma}{\mathalpha}{sfgreek}{`G}
\DeclareMathSymbol{\sDelta}{\mathalpha}{sfgreek}{`D}
\DeclareMathSymbol{\sTheta}{\mathalpha}{sfgreek}{`J}
\DeclareMathSymbol{\sLambda}{\mathalpha}{sfgreek}{`L}
\DeclareMathSymbol{\sXi}{\mathalpha}{sfgreek}{`X}
\DeclareMathSymbol{\sPi}{\mathalpha}{sfgreek}{`P}
\DeclareMathSymbol{\sSigma}{\mathalpha}{sfgreek}{`S}
\DeclareMathSymbol{\sUpsilon}{\mathalpha}{sfgreek}{`U}
\DeclareMathSymbol{\sPhi}{\mathalpha}{sfgreek}{`F}
\DeclareMathSymbol{\sPsi}{\mathalpha}{sfgreek}{`Y}
\DeclareMathSymbol{\sOmega}{\mathalpha}{sfgreek}{`W}

\DeclareRobustCommand{\mcal}[1]{%
  \ifcat\noexpand#1\relax\mathnormal{#1}\else\cal{#1}\fi
}
\DeclareRobustCommand{\BM}[1]{%
  \ifcat\noexpand#1\relax\bm{\boldUppercaseItalicGreek{#1}}\else\bm{#1}\fi
}
\makeatletter
\newcommand{\boldUppercaseItalicGreek}[1]{\csname var\expandafter\@gobble\string#1\endcsname}
\makeatother
\newcommand{\rv}[1]{\msf{#1}} 
\newcommand{\RV}[1]{\bm{\msf{#1}}}  
\newcommand{\RM}[1]{\bm{\msf{#1}}}  

\newcommand{\V}[1]{\bm{#1}} 
\newcommand{\M}[1]{\BM{#1}} 

\definecolor{BLUE}{rgb}{0,0,1}

\newtheorem{remark}{Remark}

\newcommand{\paperTitle}{Exploiting Multipath Information for Integrated Localization and Sensing via PHD Filtering}
\acrodef{gnss}[GNSS]{global navigation satellite system}
\acrodef{rf}[RF]{radio frequency}
\acrodef{aoa}[AOA]{angle-of-arrival}
\acrodef{rss}[RSS]{received signal strength}
\acrodef{toa}[TOA]{time-of-arrival}
\acrodef{tdoa}[TDOA]{time-difference-of-arrival}
\acrodef{rtt}[RTT]{round-trip time}
\acrodef{fdd}[FDD]{frequency division duplex}
\acrodef{tdd}[TDD]{time division duplex}
\acrodef{fd}[FD]{full-duplex}
\acrodef{sdp}[SDP]{semidefinite programming}
\acrodef{crlb}[CRLB]{Cram\'{e}r-Rao lower bound}
\acrodef{nc}[NC]{narrow correlator}
\acrodef{sc}[SC]{storbe correlator}
\acrodef{pll}[PLL]{phase locked loop}
\acrodef{mp}[MP]{multipath}
\acrodef{sp}[SP]{single path}
\acrodef{ff}[FF]{flat fading}
\acrodef{mds}[MDS]{multidimensional scaling}
\acrodef{snr}[SNR]{signal-to-noise ratio}
\acrodef{los}[LOS]{line-of-sight}
\acrodef{nlos}[NLOS]{non-line-of-sight}
\acrodef{sic}[SIC]{serial interference cancelation}
\acrodef{pic}[PIC]{parallel interference cancelation}
\acrodef{adc}[ADC]{analog-to-digital converter}
\acrodef{bp}[BP]{basis pursuit}
\acrodef{lasso}[LASSO]{least absolute shrinkage and selection operator}
\acrodef{omp}[OMP]{orthogonal matching pursuit}
\acrodef{lls}[LLS]{linear least squares}
\acrodef{wlls}[WLLS]{weighted linear least squares}
\acrodef{nlls}[NLLS]{nonlinear least squares}
\acrodef{awgn}[AWGN]{additive white Gaussian noise}
\acrodef{cirf}[CIRF]{channel impulse response function}
\acrodef{irf}[IRF]{impulse response function}
\acrodef{llr}[LLR]{log-likelihood ratio}
\acrodef{llrs}[LLRs]{log-likelihood ratios}
\acrodef{fim}[FIM]{Fisher information matrix}
\acrodef{efim}[EFIM]{equivalent Fisher information matrix}
\acrodef{mse}[MSE]{mean squared error}
\acrodef{peb}[PEB]{position error bound}
\acrodef{rmse}[RMSE]{root mean squared error}
\acrodef{seb}[SEB]{synchronization error bound}
\acrodef{imu}[IMU]{inertial measurement unit}
\acrodef{gm}[GM]{Gaussian-mixture}
\acrodef{reb}[REB]{ranging error bound}
\acrodef{co}[CO]{clock offset}
\acrodef{pdf}[PDF]{probability density function}%
\acrodef{iot}[IoT]{Internet-of-Things}
\acrodef{los}[LOS]{line-of-sight}
\acrodef{nlos}[NLOS]{non-line-of-sight}
\acrodef{phd}[PHD]{probability hypothesis density}
\acrodef{slam}[SLAM]{simultaneous localization and mapping}
\acrodef{mpc}[MPC]{multipath component}
\acrodef{toa}[ToA]{time-of-arrival}
\acrodef{aoa}[AoA]{angle-of-arrival}
\acrodef{vt}[VT]{virtual transmitter}
\acrodef{rbpf}[RBPF]{Rao-Blackwellised particle filter}
\acrodef{GM}{Gaussian-mixture}
\acrodef{uwb}[UWB]{ultra-wideband}
\acrodef{rfs}[RFS]{random finite set}
\acrodef{cir}[CIR]{channel impulse response}
\acrodef{ekf}[EKF]{extended Kalman filter}
\acrodef{fisst}[FISST]{finite-set statistics}
\acrodef{map}[MAP]{maximum-a-posteriori}
\acrodef{cdf}[CDF]{cumulative distribution function}
\acrodef{mmse}[MMSE]{minimum-mean-square-error}
\acrodef{xr}[XR]{extended reality}
\acrodef{gc}[GC]{Gaussian component}
\acrodef{ris}[RIS]{reconfigurable intelligent surface}
\acrodef{rhs}[RHS]{right-hand side}
\acrodef{fov}[FoV]{field of view}
\acrodef{da}[DA]{data association}
\acrodef{smc}[SMC]{sequential Monte Carlo}
\acrodef{eap}[EAP]{expected-a-posteriori}
\acrodef{ppp}[PPP]{Poisson point process}

\begin{document}

\title{\paperTitle}

\author{
    Yinuo Du,
    Hanying Zhao,~\IEEEmembership{Member,~IEEE},
    Yang Liu,
    Xinlei Yu,
    Yuan Shen,~\IEEEmembership{Senior Member,~IEEE}
    \vspace{-2.5em}
    \thanks{
         Yinuo Du, Hanying Zhao and Yuan Shen are with the Department of Electronic Engineering, and Beijing National Research Center for Information Science and Technology, Tsinghua University, Beijing 100084, China (e-mail: {duyn19@mails.tsinghua.edu.cn;
         \{hying\_zhao,shenyuan\_ee\}@tsinghua.edu.cn}). Yang Liu and Xinlei Yu are with the Department of Standards Research, OPPO Inc., Beijing 100101, China (e-mail: {liuyangbj@oppo.com;
         yuxinlei@oppo.com}).
         
         © 2024 IEEE.  Personal use of this material is permitted.  Permission from IEEE must be obtained for all other uses, in any current or future media, including reprinting/republishing this material for advertising or promotional purposes, creating new collective works, for resale or redistribution to servers or lists, or reuse of any copyrighted component of this work in other works.
         
         DOI of the published paper is 10.1109/TVT.2024.3433028
       }
}
\maketitle

\begin{abstract}
Accurate localization and perception are pivotal for enhancing the safety and reliability of vehicles. However, current localization methods suffer from reduced accuracy when the \ac{los} path is obstructed, or a combination of reflections and scatterings is present. In this paper, we present an integrated localization and sensing method that delivers superior performance in complex environments while being computationally efficient. Our method uniformly leverages various types of \acp{mpc} through the lens of \acp{rfs}, encompassing reflections, scatterings, and their combinations. This advancement eliminates the need for the multipath identification step and streamlines the filtering process by removing the necessity for distinct filters for different multipath types, a requirement that was critical in previous research. The simulation results demonstrate the superior performance of our method in  both robustness and effectiveness, particularly in complex environments where the \ac{los} path is obscured and in situations involving clutter and missed detection of \ac{mpc} measurements.
\end{abstract}

\acresetall



\acresetall		

\vspace{-1em}
\section{Introduction}\label{sec:intro}
High-accuracy localization with environmental sensibility is a critical technology in autonomous driving and 6G communication systems\cite{WinSheDai:J18}. However, achieving precise localization in complex propagation environments poses a significant challenge for vehicles, primarily due to obstructions, reflections, and signal scatterings, leading to interference and phase shifting of received signals. To bolster vehicle safety and reliability, it is essential to address these complexities in localization.  

Conventional localization methods put great efforts into eliminating these effects to extract \ac{los} signal for positioning. Nowadays, with the development of wireless technologies such as vehicle-mounted large bandwidth and large-scale antenna arrays, the high temporal- and spatial-resolution measurements enable us to separate mixed signals and estimate both the \ac{los} signal and the \ac{nlos} \acp{mpc}\cite{ZhaZhaShe:J20}. 
Therefore, there are a growing number of localization methods that aim to exploit these multipath information to improve localization accuracy and achieve environmental sensing\cite{DeiThi:C10,GenJosWanZhaDamFie:J16,KimGraGaoBatKimWym:J20,WanLiuShe:C21}. This topic is often referred to as radio \ac{slam}. We give a brief review of them and make a comparison between existing works and ours.

In \cite{DeiThi:C10}, the authors use a monostatic UWB device to detect and localize surrounding features. They adopt particle filtering with two methods of associating  measurements with landmarks: Nearest Neighbor Data Association and Probabilistic Data Association. In \cite{WanLiuShe:C21}, the authors use an adaptive federal filter to handle \ac{nlos} \acp{mpc}. However, their approach is limited to a specific number of \acp{mpc} and only considers reflection \acp{mpc}. Channel-SLAM in \cite{GenJosWanZhaDamFie:J16} treats multipath components as signals emitted from \acp{vt}, and an \ac{rbpf} based on Recursive Bayesian filtering is employed to improve computational tractability by sampling particles in a subspace.  Other works such as \cite{KimCheKesGeKeyAleKimWym:J23, YanZhaZhaDiDonYanSonLin:J22} achieve radio \ac{slam} using \acp{ris}, where Murty’s algorithm is adopted to handle \ac{da} in \cite{KimCheKesGeKeyAleKimWym:J23}. As these works have mentioned, one of the main challenges of radio-\ac{slam} lies in \ac{da}, i.e., assigning measurements to corresponding landmarks. 
In the domain of Bayesian \ac{slam}, \ac{phd}-\ac{slam} has garnered significant attention in recent years due to its superior performance compared to traditional \ac{rbpf}-\ac{slam} methods \cite{MulVoAdaVo:J11,LeuInoAda:J16}. Employing \ac{fisst}, \ac{phd}-\ac{slam} effectively addresses the complex \ac{da} problem by grouping time-varying elements into \acp{rfs}. This approach has been adopted in the context of radio-\ac{slam}\cite{AmjAhmLazHafKhanZah:J23} in works such as \cite{KimGraGaoBatKimWym:J20} to tackle the \ac{da} issue. 
The authors in  \cite{KimGraGaoBatKimWym:J20} proposed a cooperative vehicle positioning and radio environment mapping method via a multiple-model \ac{phd} filter. Although a promising performance is achieved, multiple \ac{phd} filters are derived and employed for different types of \acp{mpc} since \acp{mpc} originating from reflections and scatterings are treated distinctly. Moreover, this method fails to handle \acp{mpc} which are under the dual effect of scatterings and reflections. 

\begin{figure}[t]
	\centering
    \includegraphics[width=\columnwidth]{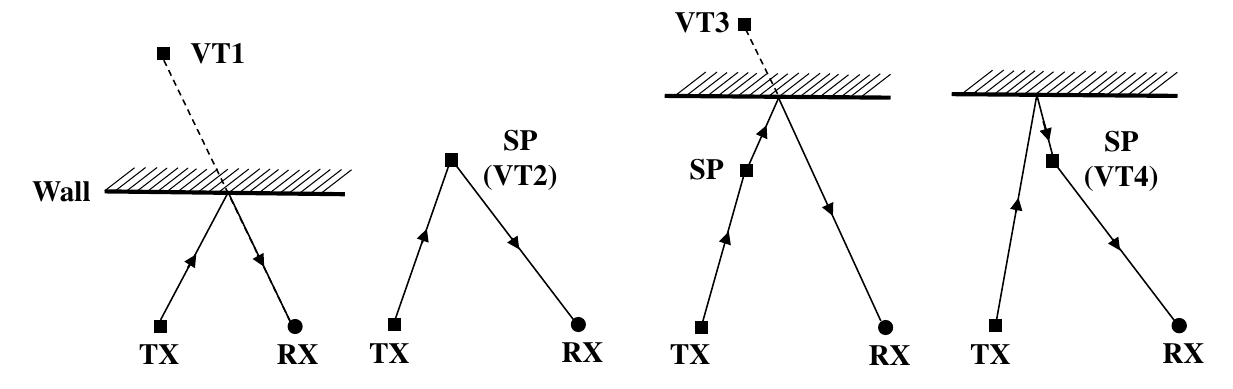}
    \vspace{-2em}
	\caption{An illustration to show the geometric relationships among transmitters (Txs), receivers (Rxs), and  \acp{vt} for various \acp{mpc}: planar reflections (left), scattering points (SPs) (second from the left), and a combination of one scattering and one reflection (third and fourth from the left). The multipath signals can be perceived as if they were virtually emitted from \acp{vt}.}
	\label{fig:system}
	\vspace{-1.5em}
\end{figure}

In this paper, we present an integrated localization and sensing method that delivers superior performance in complex environments while being computationally efficient. Our method can uniformly leverage various types of \acp{mpc}, encompassing reflections, scatterings, and their combinations, thus bypassing the multipath identification step. The key enabler is the introduced \ac{mpc} model represented as an \ac{rfs}, which augments the state space to characterize the diverse geometric relationships between the agent and the environment for each \ac{mpc} type. This advancement simplifies the filtering process by eliminating the requirement of separate filters for different types of \acp{mpc}, a step that was essential in previous studies like \cite{KimGraGaoBatKimWym:J20}. Our method offers more streamlined and versatile solutions for integrated localization and sensing in complex environments, which can serve as a practical guideline for implementation and advancement in areas such as autonomous driving and 6G.
\section{System Model}\label{sec:sysmod}
 Consider a 2D localization scenario where a vehicle (agent) receives wireless signals transmitted from a static base station (anchor) for positioning.\footnote{Our method is also applicable to  3D scenarios and multi-anchor systems. In 3D scenarios, it is adapted by adding elevation angles to the observation space and incorporating the $z$-axis into the state space. For multi-anchor systems, our approach is extended by augmenting observations from all anchors within the observation space.}  Let $\V{r}_{\rm b}=[x_{\rm b}~y_{\rm b}]^{\mathrm{T}}$ denote the position of the stationary anchor. At time instant $t_k$ with $k\in\mathbb{Z}^+$, the agent state can be described by
\begin{equation}
    \RV{x}_k=
    \begin{bmatrix}
        \RV{r}^{\mathrm T}_k &
        \RV{v}^{\mathrm T}_k &
        \rv{b}_k
    \end{bmatrix}^{\mathrm T}
    \label{agentstate}
\end{equation}
where  $\RV{r}_k=[\rv{x}_k~ \rv{y}_k]^{\mathrm{T}}$,  $\RV{v}_k=[\rv{v}_{x_k}~ \rv{v}_{y_k}]^{\mathrm{T}}$ and $\rv{b}_k$ denote position, velocity and ranging bias incurred by clock offset, respectively. At time $t_{k+1}:=t_{k}+\Delta t$, the agent new state yields:
{\setlength{\abovedisplayskip}{3pt}
\setlength{\belowdisplayskip}{3pt}
\begin{equation}
    \label{eq:motionmodel}
    \begin{split}
        \RV{x}_{k+1}=
        \M{A}
        \RV{x}_{k}
        +
        \M{B}
        \RV{n}_{k}
    \end{split}
\end{equation}
where  $\RV{n}_k = [\rv{u}_{{\rm x},k}~\rv{u}_{{\rm y},k}~\rv{u}_{{\rm b},k}]^{
\mathrm T}$ is the process noise modelled as \ac{awgn}  with variances $\sigma_{\rm x}^2$, $\sigma_{\rm y}^2$ and $\sigma_{\rm b}^2$. $\M{A}$ and  $\M{B}$ are transition matrices, which can be obtained from 
$\RV{r}_{k+1}=\RV{r}_k+\RV{v}_k\Delta t+[\rv{u}_{x,k}~\rv{u}_{y,k}]^{
\mathrm T}\Delta t^2/2$, $\RV{v}_{k+1}=\RV{v}_k+[\rv{u}_{x,k}~\rv{u}_{y,k}]^{\mathrm T}\Delta t$,  and $\rv{b}_{k+1}=\rv{b}_k+\rv{u}_{{\rm b},k}\Delta t$. }

\subsection{Multipath Model}\label{sec:mulmod}
In complicated environments, where reflection surfaces (like exterior walls of buildings) and scatterers (like smaller objects along the road) are common, wireless signals often reflect and scatter before reaching the agent via multiple paths.  As \ac{nlos} \acp{mpc} embed information pertinent to both agent localization and environmental context, we exploit those \acp{mpc} to enhance localization accuracy and achieve environment sensing.

In general, \ac{nlos} \acp{mpc} can be categorized into three types, as shown in Fig. \ref{fig:system}: 1) \acp{mpc} induced by reflections; 2) \acp{mpc} caused by scatterings; and 3) \acp{mpc} influenced by both scatterings and reflections. The diverse geometric relationships between the agent and the environment for each \ac{mpc} type present a significant challenge: identifying these heterogeneous multipaths. This involves determining the specific category to which each \ac{mpc} belongs. To overcome this deficiency, we introduce a special form of \acfp{vt}, allowing us to treat all types of \acp{mpc} in a unified way. Through the lens of \ac{vt}, all \ac{nlos} \acp{mpc} can be modeled as if they are virtually emitted from \acp{vt} \cite{GenJosWanZhaDamFie:J16} and described as
{\setlength{\abovedisplayskip}{3pt}
\setlength{\belowdisplayskip}{3pt}
\begin{equation}\label{eq:vt_model}
    \V{m}^{(l)}=
    \begin{bmatrix}
        (\V{r}^{(l)}_{\rm vt})^\mathrm{T} & b^{(l)}_{\rm vt}
    \end{bmatrix}^{\mathrm T}
\end{equation}
where  $\V{r}^{(l)}_{\rm vt} = [x^{(l)}_{\rm vt}~ y^{(l)}_{\rm vt}]^{\mathrm{T}}\in \mathbb{R}^2 $ denote the 2-D position of the $l$-th \ac{vt} and $b^{(l)}_{\rm vt}$ is the additional propagation length of this \ac{mpc}.}
Taking the four \acp{mpc} in Fig. \ref{fig:system} for illustration:
\begin{enumerate}
    \item The reflection \ac{mpc} can be viewed as being directly transmitted from the \ac{vt} which is the anchor's mirror image to the reflector (VT1);
    \item The \Ac{mpc} induced by a scatterer can be viewed as being directly transmitted from the scatterer $\V{r}^{(2)}_{\rm vt} :=\V{r}_{\rm S}=[x_{\rm S}~y_{\rm S}]^{\mathrm{T}}$ with an additional propagation bias $b^{(2)}_{{\rm vt}} =\|\V{r}_{\rm b}-\V{r}_{S}\|_2$ (VT2);
    \item The \acp{mpc} under the dual effect of scattering and reflection can also be viewed as being directly transmitted from a \ac{vt} with an additional propagation bias. As in Fig. \ref{fig:system}, \acp{mpc} of VT3 and VT4 are composed of one reflection and one scattering each, albeit in a different sequence. VT3 is located at the scatterer's mirror image to the reflector with $b^{(3)}_{{\rm vt}} = \|\V{r}_{\rm b}-\V{r}_{S}\|_2$ and VT4 is located  at the same location as the scatterer with $b^{(4)}_{{\rm vt}} = \|\V{r}_{\rm b}-\V{r}^{(3)}_{\rm vt}\|_2$.
\end{enumerate}

\subsection{Formulating Multipath Components Using \ac{rfs}}\label{sec:wirmeas}
 As \acp{mpc} emerge and die along with the movement of the agent, and due to clutter and the missed detection phenomena, we need to identify which \acp{vt} are present and more importantly, manage the NP-hard \ac{da} task for both \acp{vt} and observations across various time instances. To address this issue, we group \acp{vt} and wireless measurements into \acp{rfs}\cite{Mah:B07} instead of the conventional matrices. 
 By adopting the \ac{rfs} framework, we can more adeptly characterize the stochastic nature of both the detection process and the environment's dynamics for radio-\ac{slam}. We formulate the \ac{vt} map, which encompasses all detected \acp{vt}, as an \ac{rfs} to accommodate its time-varying characteristics. This set  at time instant $t_k$ is denoted by $\mathcal{M}_k=\{\RV{m}_k^{(1)}, \RV{m}_k^{(2)}, ...,\RV{m}_k^{(\left|\mathcal{M}_k\right|)}\}$.

\begin{remark}
The benefits of this \ac{vt} model are summarized as follows: First, our method enables a consistent description and utilization of all types of \acp{mpc}, including those arising from multiple reflections and scatterings. In contrast, other approaches use distinct models for different types of \acp{mpc}, necessitating the identification of the \ac{mpc} type prior to filtering or the usage of multiple filters in the filtering stage. Moreover, existing methods are unable to tackle \acp{mpc} affected by the combined effects of reflections and scatterings. 
 Second, a challenge encountered in previous works is the difficulty to distinguish \acp{mpc} whose \acp{vt} are located at the same spatial position. This situation often arises when different \acp{mpc}  undergo the same scatterer before reaching the receiver, such as VT2 and VT4 in Fig. \Ref{fig:system}. The introduced \ac{vt} model overcomes this challenge by including the additional propagation bias parameter, as shown in \eqref{eq:vt_model}, which is distinct for VT2 and VT4, allowing for their differentiation. This feature facilitates the development of the method for accurately discerning environmental characteristics. 
 Third, utilizing \acp{rfs}, we explicitly bypass the NP-hard \ac{da} step. 
 These advantages highlight the effectiveness and practicality of our \ac{vt} model in handling complex propagation environments and leveraging multipath information for improved localization and sensing. 
\end{remark}

 On the other hand, the \acp{toa} and \acp{aoa} of detected \acp{mpc} are estimated from the \ac{cir} \cite{WanLiuShe:C21}  \cite{ZhaHuaShe:J22}. Then, the entire set of measurements at time instant $t_k$ under the \ac{rfs} framework can be described by
\begin{equation}
    \label{eq:measurementset}
    \mathcal{Z}_{k}=\{ \mathcal{Z}^{(0)}_{k},\mathcal{Z}^{\rm NLOS}_{k}\}
\end{equation}
where \acp{rfs} $\mathcal{Z}^{(0)}_{k}$ and $\mathcal{Z}^{\rm NLOS}_{k}$ describe the \ac{los} and the \ac{nlos} measurements at time instant $t_k$, respectively, which are non-empty when \ac{los} signal/\ac{nlos} \acp{mpc} are detected or false alarms occur, and are empty otherwise, i.e.
\begin{equation*}
    \mathcal{Z}^{(0)}_{k}=\begin{cases}
    \{\RV{z}^{(0)}_{k}\}, & \text{\ac{los} signal detected,}   \\
    \emptyset, & \text{\ac{los} signal miss,}
    \end{cases}~\mathcal{Z}^{\rm NLOS}_{k}=\bigcup_{l\ge 1}\{\RV{z}_{k}^{({l})}\}
\end{equation*}
and the detection probability for a \ac{vt} located at $\V{m}$ with the agent located at $\V{x}$ is denoted as $P_{\rm D}(\V{m},\V{x})$. The range-bearing observation of the $l$-th \ac{vt} $\RV{m}_k^{(l)}$  is 
\begin{align}\label{eq:nlosmeasurement}
    \begin{split}
    \RV{z}_{k}^{({l})}&=
    \begin{bmatrix}
        \rv{d}_{k}^{({l})} \\[0.7em]
        \rv{\theta}_{k}^{({l})}
    \end{bmatrix} =
    \begin{bmatrix}
        \|\RV{r}_{k}-\RV{r}^{(l)}_{\rm vt}\|_2+\rv{b}_k+\rv{b}^{(l)}_{\rm vt}\\
        \arctan\left(\frac{\rv{y}^{(l)}_{{\rm vt}}-\rv{y}_k}{\rv{x}^{(l)}_{{\rm vt}}-\rv{x}_k}\right)
    \end{bmatrix}
    +
    \RV{\omega}_{k}^{(l)}
    \end{split}
\end{align}
in which $\RV{\omega}_{k}^{(l)}$ denotes the \ac{awgn} with standard deviation $\sigma_{\rm d}$ and $\sigma_{\rm \theta}$ for the range and the bearing measurements, respectively. The \ac{los} measurement $\RV{z}^{(0)}_{k}$ has the same structure as that in \eqref{eq:nlosmeasurement} but with noise standard deviations $\sigma^{(0)}_{\rm d}$ and $\sigma^{(0)}_{\theta}$. All the measurements at time instance $t_k$, including real measurements and clutter are grouped in \eqref{eq:measurementset}. 

To achieve integrated localization and sensing, we aim to simultaneously estimate the agent trajectory $\RV{x}_{0:k}$ and the \ac{vt} map $\mathcal{M}_k$ based on $\mathcal{Z}_{1:k}$, representing measurements from $t_1$ to $t_k$, i.e., computing the posterior distribution {$p(\RV{x}_{0:k},\mathcal{M}_k|\mathcal{Z}_{1:k})$}.
\begin{remark}
A potential limitation of our \ac{vt} model is its requirement for multi-shot measurements. This is because the \ac{vt}'s state is characterized by a 3-D vector, whereas the range-bearing measurements at a one time stamp are 2-D. To let the volume of observations surpass the number of unknowns, we leverage multi-shot measurements and determine the agent trajectory alongside the states of \acp{vt}  through filtering.
\end{remark}
\section{Integrated Localization and Environment Sensing via PHD Filtering}\label{sec:algo}

\begin{figure}[t]
    \centering
    \includegraphics[width=0.9\columnwidth]{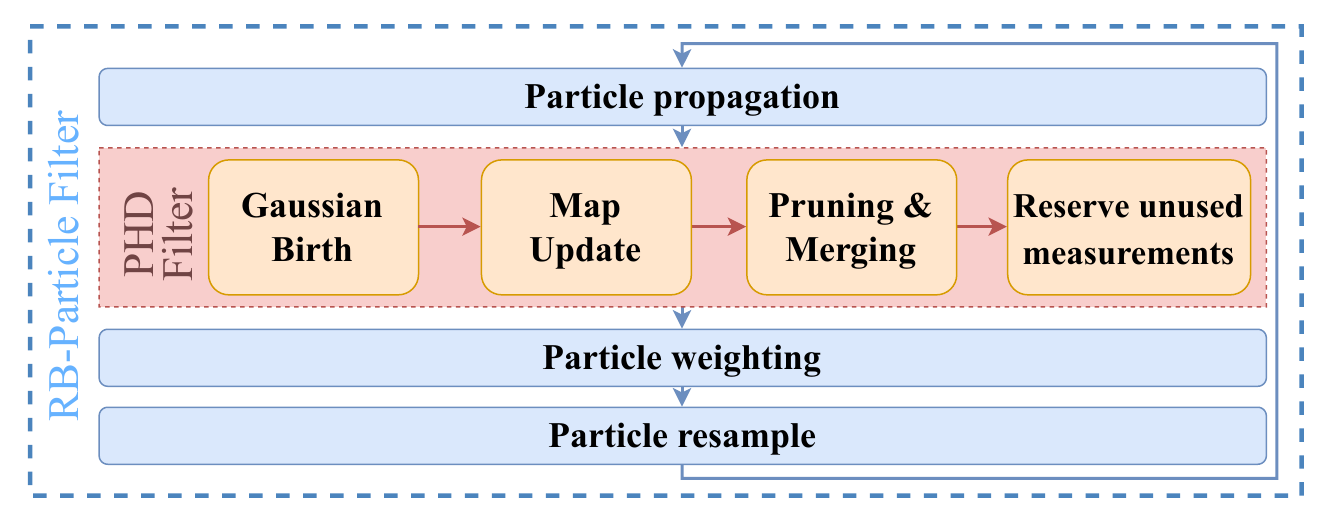}
	\caption{A General Look at the complete PHD-SLAM algorithm}
	\label{fig:filtermodel}
 \vspace{-1.5em}
\end{figure}

Our method for estimating the agent trajectory and \ac{vt} map includes four steps: particle propagation, map prediction, map update and particle reweight/resampling, as shown in Fig. \ref{fig:filtermodel}.

{\setlength{\abovedisplayskip}{3pt}
\setlength{\belowdisplayskip}{3pt}
 The computation of $p(\RV{x}_{0:k},\mathcal{M}_k|\mathcal{Z}_{1:k})$ using a homogeneous filter is challenging because $\RV{x}_k$ is a vector and $\mathcal{M}_k$ is an \ac{rfs}. To address this problem, we decompose the posterior density based on the Rao-Blackwellisation step \cite{MurRus:B01} in \acp{rbpf}
\begin{equation}
    \label{eq:decompos}
    p(\RV{x}_{0:k},\mathcal{M}_k|\mathcal{Z}_{1:k})
    =
    p(\RV{x}_{0:k}|\mathcal{Z}_{1:k})
    p(\mathcal{M}_k|\mathcal{Z}_{1:k},\RV{x}_{0:k})
\end{equation}
where the first term on the \ac{rhs} is the marginal posterior probability distribution of the agent trajectory $\RV{x}_{0:k}$, and the second term is the conditional probability distribution of the \ac{vt} map $\mathcal{M}_k$. 
{Leveraging this decomposition, similar to the approach in \cite{MulVoAdaVo:J11}, we construct an \ac{rbpf}\cite{MurRus:B01}. For a standard \ac{rbpf}, the marginal distribution  $p(\RV{x}_{0:k}|\mathcal{Z}_{1:k})$ is sampled using particles. For each particle, the conditional map $p(\mathcal{M}_k|\mathcal{Z}_{1:k},\RV{x}_{0:k})$ should be iterated analytically\cite{MurRus:B01}. Calculating and iterating the precise probability distribution of an \ac{rfs} is computational intractable, thus we use a \ac{ppp} to approximate the \ac{rfs} by aligning its intensity function with the first-order statistical moment of the \ac{rfs}, known as the \acf{phd}. This approach forms a \ac{phd} filter \cite{MulVoAdaVo:J11,LeuInoAda:J16}, details on the \ac{phd} of an \ac{rfs} and the \ac{phd} filter can be found in \cite{Mah:B07}.
}
}

\subsection{Importance Sampling and Particle Propagation}
At time instant $t_k$, $N$ particles are used to describe the agent trajectory $\RV{x}_{0:k}$, where the posterior weight of the $i$-th particle with trajectory $\V{x}^{(i)}_{0:k}$ is denoted as $w^{(i)}_{k|k}$. 
{
For the $i$-th particle, we use $v^{(i)}_{k|k}(\V{m})$ to describe the \ac{phd} of the posterior \ac{vt} map \ac{rfs} $\mathcal{M}_k|\mathcal{Z}_{1:k},\V{x}^{(i)}_{0:k}$ at $\V{m}$.}
Therefore, the posterior particle set at time instant $t_k$ is:
{\setlength{\abovedisplayskip}{3pt}
\setlength{\belowdisplayskip}{3pt}
\begin{equation}
    \mathcal{P}_k=\{\V{x}^{(i)}_{0:k},v^{(i)}_{k|k}(\V{m}),w^{(i)}_{k|k}\}_{i=1}^{N}.
    \label{eq:particleset}
\end{equation}
The posterior \ac{pdf} of the agent {trajectory} is approximated as:
$
    p(\RV{x}_{0:k}|\mathcal{Z}_{1:k})=\sum_{i=1}^{N}w^{(i)}_{k|k}\delta(\V{x}_{0:k}-\V{x}^{(i)}_{0:k})
$.
{
The conditional map distribution can be analytically approximated by its \ac{phd} as follows\cite{Mah:B07}:
\begin{equation}\label{eq:pppapproximation}
    p(\mathcal{M}_k|\mathcal{Z}_{1:k},\V{x}^{(i)}_{0:k})=e^{-\int v^{(i)}_{k|k}(\V{s})\rm d\V{s}}\prod_{\V{m}^{(l)}_k\in \mathcal{M}_k}v^{(i)}_{k|k}(\V{m}^{(l)}_k).
\end{equation}
}
}

{\setlength{\abovedisplayskip}{3pt}
\setlength{\belowdisplayskip}{3pt}
In the particle propagation step at $t_k$, we estimate the predicted distribution $p(\RV{x}_{0:k}|\mathcal{Z}_{1:k-1})$ using $\mathcal{P}_{k-1}$. Particle trajectory $\V{x}^{(i)}_{0:k}$ and the weight $w^{(i)}_{k|k-1}$ are predicted based on the principle of importance sampling\cite{AruMasGorCla:J02}. We use the importance density $q(\RV{x}^{(i)}_{0:k}|\mathcal{Z}_{1:k})$ to sample particle states, which satisfies $
    q(\RV{x}^{(i)}_{0:k}|\mathcal{Z}_{1:k})=q(\RV{x}^{(i)}_{k}|\RV{x}^{(i)}_{0:k-1},\mathcal{Z}_{1:k})q(\RV{x}^{(i)}_{0:k-1}|\mathcal{Z}_{1:k-1})
$ and calculate predicted particle weight $w^{(i)}_{k|k-1}$ by
\begin{equation}
    w^{(i)}_{k|k-1}\propto w^{(i)}_{k-1|k-1}\frac{p(\RV{x}^{(i)}_{k}|\RV{x}^{(i)}_{k-1})}{q(\RV{x}^{(i)}_{k}|\RV{x}^{(i)}_{0:k-1},\mathcal{Z}_{1:k})}.
    \label{eq:weightpredict}
\end{equation}
{Similar to the widely adopted approach in \cite{KimGraGaoBatKimWym:J20,MulVoAdaVo:J11,LeuInoAda:J16}, we use the motion model to characterise the importance density.} Specifically, we use  $q(\RV{x}^{(i)}_{k}|\RV{x}^{(i)}_{0:k-1},\mathcal{Z}_{1:k})=p(\RV{x}^{(i)}_{k}|\RV{x}^{(i)}_{k-1})$ to simplify calculations, which yields $w^{(i)}_{k|k-1}=w^{(i)}_{k-1|k-1}$.}

\subsection{\ac{phd} Filtering}
Next, we address each particle's \ac{vt} map individually. For the posterior \ac{phd} of the \ac{vt} map for the $i$-th particle at time $t_k$, we model it using a \ac{GM} form{\cite{MulVoAdaVo:J11}}:
{\setlength{\abovedisplayskip}{3pt}
\setlength{\belowdisplayskip}{3pt}
\begin{equation}
    v^{(i)}_{k|k}(\V{m})=\sum_{j=1}^{M^{(i)}_{k|k}}\alpha^{(i)}_{k|k,j}\mathcal{N}(\V{m};\V{\mu}^{(i)}_{k|k,j},\M{C}^{(i)}_{k|k,j})
    \label{eq:gmphdcomposition}
\end{equation}
where $M^{(i)}_{k|k}$ is the number of \acp{gc} in the \ac{GM}. After the particle propagation at time stamp $t_k$, only $v^{(i)}_{k-1|k-1}(\V{m})$ is accessible. To compute $v^{(i)}_{k|k}(\V{m})$, we iterate through two steps: prediction and update.}

\subsubsection{\ac{phd} Prediction}\label{sec:phd_predict}
In complex propagation environments, \acp{mpc} emerge and die along with the movement of the agent. To effectively process changeable \acp{vt}, we adopt the ``adaptive birth strategy" \cite{HouJerLan:C10}, where the mean and variance of birth \acp{gc} are determined by measurements, rather than using a fixed birth \ac{gc} distribution. This strategy has been widely employed in \ac{gm}-\ac{phd} filters and \ac{phd}-\ac{slam} works, such as \cite{KimGraGaoBatKimWym:J20,LeuInoAda:J16}. Here, to generate a 3-D \ac{gc} based on 2-D range-bear measurements at a single stamp, we derive a new method to generate birth \acp{gc}, which is tailored specifically for our \ac{vt} model.

 As shown in Fig. \ref{fig:birthGC}, for particle $i$, suppose there are $B^{(i)}_k$ \ac{mpc} measurements that do not match with any existing \acp{vt} at time instant $t_{k-1}$. To simplify the notations, suppose that the $l$-th unmatched \ac{mpc} measurement has the value of $\RV{z}_{{\rm B}_k,l}=[\rv{d},\rv{\theta}]^{\mathrm{T}}$, we generate a birth \ac{gc} with respect to this measurement, taking the form:
\begin{align*}
   v^{(i)}_{{\rm B}_k, l}(\V{m}) = \mathcal{N}(\V{m};\RV{\mu}_{{\rm B}_k,l}^{(i)},\RM{C}_{{\rm B}_k,l}^{(i)})
\end{align*}
and initialize the mean $\RV{\mu}_{{\rm B}_k,l}^{(i)} =[(\RV{r}_{{\rm B}_k,l}^{(i)})^{\mathrm{T}}~ \rv{b}_{{\rm B}_k,l}^{(i)}]^{\mathrm{T}}$ and covariance $\RM{C}_{{\rm B}_k,l}^{(i)}$ by the measurement $\RV{z}$ as follows:

\begin{figure}[t]
	\centering
	\vspace{-1.5em}
    \includegraphics[width=0.65\columnwidth]{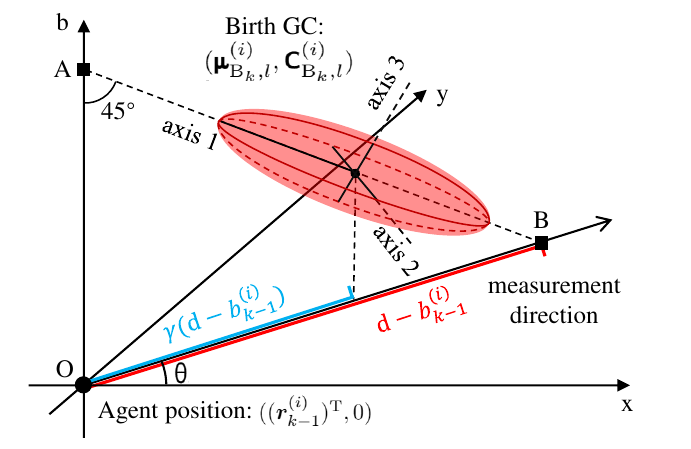}
    \vspace{-0.5em}
    \caption{For the $i$-th particle with state $\V{x}_k^{(i)}$ at time instant $t_k$, parameters of the birth \ac{gc} $(\RV{\mu}_{{\rm B}_k,l}^{(i)},\RM{C}_{{\rm B}_k,l}^{(i)})$ is generated according to the measurement $\RV{z}_{{\rm B}_k,l}=[\rv{d},\rv{\theta}]^{\mathrm{T}}$ and the particle state $\V{x}_{k-1}^{(i)}$. $\RV{\mu}_{{\rm B}_k,l}^{(i)}$ is constrained on the line AB. axis 1,2,3 are the three main axes of the \ac{gc}'s confidence ellipsoid, where axis 1 is set along line AB and axis 3 is set within plane OAB.}
    \label{fig:birthGC}
    \vspace{-1em}
\end{figure}

First, in case of the birth \ac{gc}'s mean $\RV{\mu}_{{\rm B}_k,l}^{(i)}$, with angle measurement $\rv{\theta}$, we set $\RV{r}_{{\rm B}_k,l}^{(i)}$ to the same direction relative to the agent state as $\rv{\theta}$. With range measurement $\rv{d}$, we have: 
\begin{equation}  \label{eq:toaconstraint}
    \|\V{r}^{(i)}_{k-1}-\RV{r}_{{\rm B}_k,l}^{(i)}\|_2+ \rv{b}_{{\rm B}_k,l}^{(i)} + {b}^{(i)}_{k-1}= \rv{d}
\end{equation}
due to the unknown additional propagation bias $\rv{b}_{{\rm B}j,k}^{(i)}$, $\RV{r}_{{\rm B}k,j}^{(i)}$ still has ambiguity, to address this issue, we introduce parameter $\gamma$ $(0\le\gamma\le1)$, where
\begin{equation}
    \|\V{r}^{(i)}_{k-1}-\RV{r}_{{\rm B}_k,l}^{(i)}\|_2=\gamma(\rv{d}-{b}^{(i)}_{k-1})
    \label{eq:birthconstraint}
\end{equation}
which allows us to control the mean of birth \ac{gc} by $\gamma$. 

{\setlength{\abovedisplayskip}{3pt}
\setlength{\belowdisplayskip}{3pt}
Second, in case of the \ac{gc}'s covariance $\RM{C}_{{\rm B}_k,l}^{(i)}$, we set:
\begin{equation}
    \RM{C}_{{\rm B}_k,l}^{(i)}=
    \M{T}(\rv{\theta})
    \M{\Sigma}(\rv{d})
    \M{T}(\rv{\theta})^{\mathrm{T}}
    \label{eq:birthcov}
\end{equation}
where $\M{T}({\theta})$ is a three-dimension rotational matrix, given by
\begin{equation}
    \M{T}({\theta})=        
    \begin{bmatrix}
        \frac{\cos({\theta})}{\sqrt{2}} & -\sin({\theta}) & \frac{\cos({\theta})}{\sqrt{2}}\\[0.7em]
        \frac{\sin({\theta})}{\sqrt{2}} & \cos({\theta}) & \frac{\sin({\theta})}{\sqrt{2}}\\[0.7em]
        -\frac{1}{\sqrt{2}} & 0 & \frac{1}{\sqrt{2}}\\
    \end{bmatrix}
\end{equation}
and $\M{\Sigma}(d)$ is a diagonal matrix depended on distance $d$ by $\left[\M{\Sigma}(d)\right]_{11}=\zeta\cdot (d-{b}^{(i)}_{k-1})^2 $, $\left[\M{\Sigma}(d)\right]_{22}=\iota\cdot(d-{b}^{(i)}_{k-1})^2(\sigma_{\rm \theta})^2 $ and $\left[\M{\Sigma}(d)\right]_{33}=\xi\cdot(\sigma_{\rm d})^2$ with user-defined non-negative parameters $\zeta, \iota$, and $\xi$. $\left[\M{\Sigma}(d)\right]_{11},\left[\M{\Sigma}(d)\right]_{22},\left[\M{\Sigma}(d)\right]_{33}$ denote the variance of the birth \ac{gc} along axis 1,2 and 3 in Fig. \ref{fig:birthGC} respectively.}

Then, we predict the map \ac{phd} based on the summation of \acp{gm} \cite{MulVoAdaVo:J11} using the posterior map \ac{phd} at $t_{k-1}$ and birth \acp{gc} as follows:
{\setlength{\abovedisplayskip}{3pt}
\setlength{\belowdisplayskip}{3pt}
\begin{align}
     v^{(i)}_{k|k-1}(&\V{m})  = v^{(i)}_{k-1|k-1}(\V{m})+\sum_{l=1}^{B^{(i)}_k}\alpha_{\rm B}\mathcal{N}(\V{m};\V{\mu}_{{\rm B}_k,l}^{(i)},\M{C}_{{\rm B}_k,l}^{(i)}) \notag\\
        &= \sum_{j=1}^{M^{(i)}_{k|k-1}}\alpha^{(i)}_{k|k-1,j}\mathcal{N}(\V{m};\V{\mu}^{(i)}_{k|k-1,j},\M{C}^{(i)}_{k|k-1,j})
    \label{eq:phdpredict}
\end{align}
where $M^{(i)}_{k|k-1}=M^{(i)}_{k-1|k-1}+B^{(i)}_k$ and $\alpha_{\rm B}$ denotes the \ac{gc}'s birth weight. }

\subsubsection{\ac{phd} Update}
With  wireless measurements at time $t_{k}$, we then update the \ac{phd} map of each particle. For particle $i$, denoting $P_{{\rm D}}(\V{\V{\mu}^{(i)}_{k|k-1,j}},\V{x}^{(i)}_{k})$ as $P_{{\rm D},k,j}^{(i)}$, the posterior \ac{phd} is updated using the GM-PHD filter's corrector equation\cite[Sec. 6.D]{VoMalBarCorOsbMahVo:B15}:
{\setlength{\abovedisplayskip}{3pt}
\setlength{\belowdisplayskip}{3pt}
\begin{align}
&v^{(i)}_{k|k}(\V{m})=
\sum_{l=1}^{\mid \mathcal{Z}_{k}^{{\rm NLOS}} \mid}
\sum_{j=1}^{M^{(i)}_{k|k-1}}\alpha^{(i)}_{k,l,j}  \mathcal{N}(\V{m};\V{\mu}^{(i)}_{k,l,j},\M{C}^{(i)}_{k,l,j})
\\
&+\sum_{j=1}^{M^{(i)}_{k|k-1}}(1-P_{{\rm D},k,j}^{(i)})\alpha^{(i)}_{k|k-1,j}\mathcal{N}(\V{m};\V{\mu}^{(i)}_{k|k-1,j},\M{C}^{(i)}_{k|k-1,j})\notag
\end{align}
where the second term is induced by missed detections. For the first term, $\V{\mu}^{(i)}_{k,l,j}$ and $\M{C}^{(i)}_{k,l,j}$ are the mean and variance of the $j$-th \ac{gc} corrected by the $l$-th measurement in $\mathcal{Z}_{k}^{\rm NLOS}$ using \acp{ekf}\cite{MulVoAdaVo:J11}. Moreover, we model clutter as a Poisson point process with intensity $\kappa(\V{z})$ and the expected number of clutter as $\lambda_{\rm c}$. The corrected \acp{gc}' weights can be updated like a common \ac{phd} filter as in \cite{MulVoAdaVo:J11,LeuInoAda:J16}:
\begin{equation}
    \alpha^{(i)}_{k,l,j}=
    \frac{P_{{\rm D},k,j}^{(i)}\alpha_{k|k-1,j}^{(i)}q^{(i)}_{k,j}(\RV{z}_k^{(l)})}
    {\kappa({\RV{z}_k^{(l)}})+\sum_{s=1}^{M^{(i)}_{k|k-1}}P_{{\rm D},k,j}^{(i)}\alpha_{k|k-1,s}^{(i)}q^{(i)}_{k,s}(\RV{z}_k^{(l)})}
\end{equation}
where $q^{(i)}_{k,j}(\RV{z}_k^{(l)})$ is the probability density of the $l$-th measurement raised by the $j$-th \ac{vt} \ac{gc}\cite{MulVoAdaVo:J11}. 
In the filtering process, coarse gating\cite{VoMalBarCorOsbMahVo:B15} is employed to ease computational burden and determine which \ac{mpc} measurements are reserved for prediction in the next time stamp.}

\begin{figure*}[t]
    
	\centering
	\subfigure[Scenario 1: VT1 with 1 reflection, VT2 with 1 scattering, VT3 and VT4 with 1 reflection and 1 scattering.]{
        \includegraphics[width=0.3\textwidth]{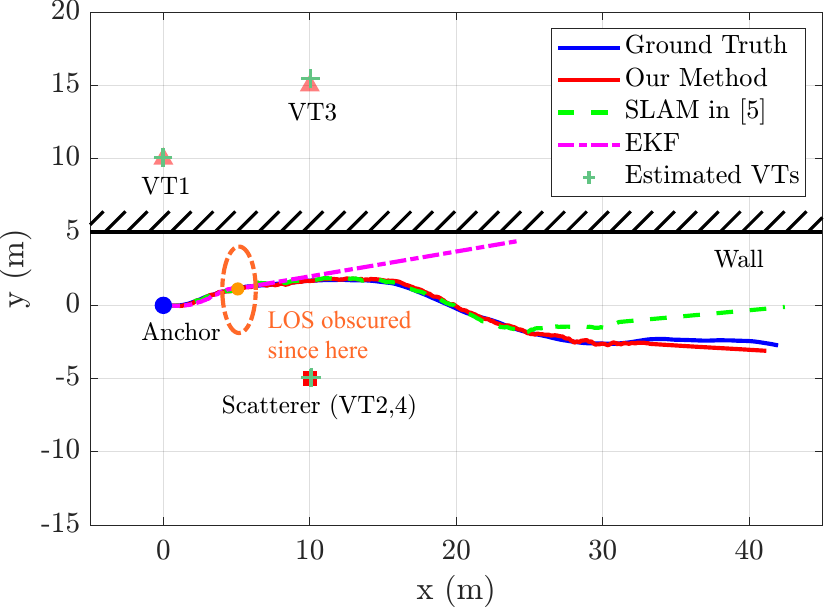}
        \label{fig:simtraj1}
    }
	\subfigure[Scenario 2: VT1, VT2 with 1 reflection, VT3 with 1 scattering, VT4, VT5 with 2 reflections, VT6,7,8,9 with 1 reflection and 1 scattering.]{
        \includegraphics[width=0.3\textwidth]{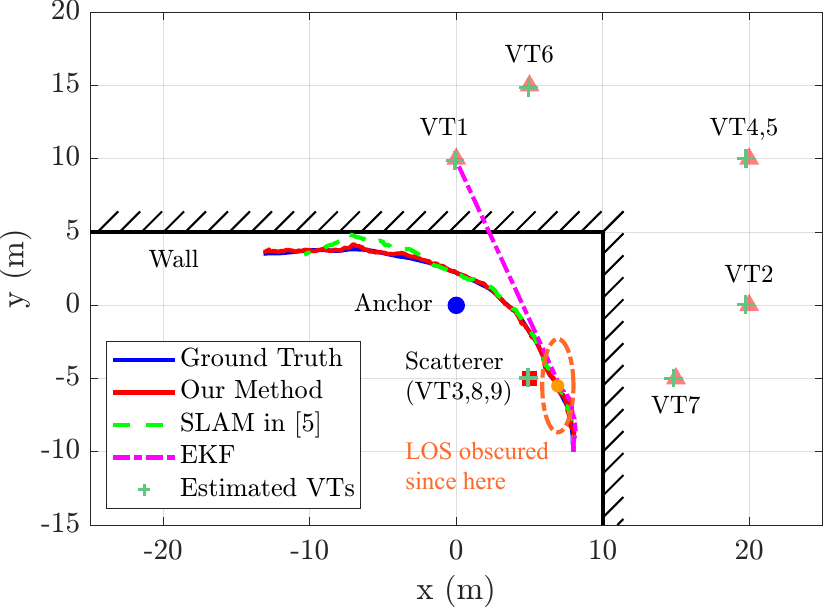}
        \label{fig:simtraj2}
    }
	\subfigure[The \ac{cdf} of localization errors of three methods in two scenarios.]{
        \includegraphics[width=0.3\textwidth]{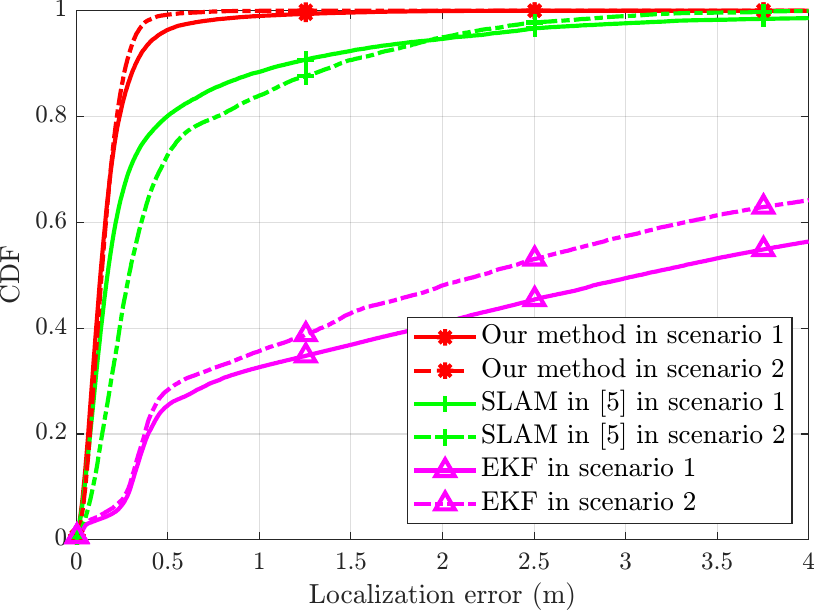}
        \label{fig:simcdf}
    }
    \vspace{-0.5em}
    \caption{(a) and (b) illustrate the configurations of 2 simulation scenarios and the localization results of one representative experiment of three methods for each scenario, (c) shows the \ac{cdf} of localization errors.}
    \vspace{-1.5em}
    \label{fig:simresult}
\end{figure*}

\subsection{Particle Reweight and Resample}
Combing the updated \ac{vt} map and agent state, we have:
{\setlength{\abovedisplayskip}{3pt}
\setlength{\belowdisplayskip}{3pt}
{
\begin{equation}
\begin{split}
    p(\RV{x}_{0:k}|\mathcal{Z}_{1:k})&\propto p(\RV{x}_{0:k},\mathcal{Z}_{k}|\mathcal{Z}_{1:k-1})\\
    &= p(\mathcal{Z}_{k}|\RV{x}_{0:k},\mathcal{Z}_{1:k-1})p(\RV{x}_{0:k}|\mathcal{Z}_{1:k-1}).
\end{split}
\label{eq:improtanceweighting}
\end{equation}}}
The updated particle weight for particle $i$ is thus given by
{
\begin{equation}
    \begin{split}
    w^{(i)}_{k|k}&\propto
    p(\mathcal{Z}_{k}|\V{x}^{(i)}_{0:k},\mathcal{Z}_{1:k-1})w^{(i)}_{k|k-1}  \\
    &= p(\mathcal{Z}_{k}^{(0)}|\V{x}^{(i)}_{k})p(\mathcal{Z}_{k}^{\rm NLOS}|\V{x}^{(i)}_{0:k},\mathcal{Z}_{1:k-1})w^{(i)}_{k|k-1}
    \end{split}
    \label{eq:weightcompute}
\end{equation}
}
where the first term is the contribution of \ac{los} measurement and is set equal for all particles if \ac{los} measurement is lost. As shown in \cite{LeuInoAda:J16}, the second term comes from the \ac{nlos} \acp{mpc} and satisfies
{\begin{align}\label{eq:multifeature}
      p(\mathcal{Z}_{k}^{\rm NLOS}&|\V{x}^{(i)}_{0:k},\mathcal{Z}_{1:k-1})= \\
    &p(\mathcal{Z}_{k}^{\rm NLOS}|\V{x}^{(i)}_{0:k},\Lambda_{k})\frac{p(\Lambda_{k}|\V{x}^{(i)}_{0:k},\mathcal{Z}_{1:k-1})}{p(\Lambda_{k}|\V{x}^{(i)}_{0:k},\mathcal{Z}_{1:k})} \notag
\end{align}}
where $\Lambda_{k}$ is a free map variable. Multiple strategies exist considering the choice of $\Lambda_{k}$, including empty-set strategy, single-feature strategy and multi-feature strategy\cite{LeuInoAda:J16}. To generate more accurate estimates, we adopt the multi-feature strategy to decide $\Lambda_{k}$, as shown in \cite{LeuInoAda:J16}.
When the  number of effective particles is low, particle resampling will be executed.

Finally,  the agent trajectory is estimated based on the \ac{mmse} criteria using a weighted mean approach, and the \ac{vt} map is estimated by extracting \acp{gc} with highest weights as demonstrated in previous works such as \cite{LeuInoAda:J16,MulVoAdaVo:J11}.

\section{Numerical Results}\label{sec:numerical}
In this section, two distinct simulation scenarios are investigated, with the layouts of their respective environments illustrated in Fig. \ref{fig:simtraj1} and Fig. \ref{fig:simtraj2}. 
In the first scenario, shown in \ref{fig:simtraj1}, there are a base station, a point scatterer, and a planar wall, located at $(0,0) \rm m$, $(10,-5) \rm m$, and along $y=10 \rm m$, respectively. This configuration will lead to four \ac{nlos} \acp{mpc} and four corresponding \acp{vt}.\footnote{While our method can handle \acp{mpc} with any combination of reflections and scatterings, in our simulations, we only consider \ac{nlos} \acp{mpc} related with either up to two reflections, two scatterings, or a combination of one reflection and one scattering.} {The second scenario, depicted in Fig. \ref{fig:simtraj2}, presents a more complex propagation environment. 
It includes a base station located at $(0,0) \rm m$, a point scatterer at $(5,-5) \rm m$, a horizontal wall along $y=5 \rm m$, and a vertical wall along $x=10 \rm m$.
This setup will result in 9 \acp{vt}. We also investigate the agent's \ac{fov}, which contributes to a more realistic representation of the real-life scenario. \acp{vt} located beyond the agent's \ac{fov} distance are not detected, aligning with practical limitations.}

In both simulation scenarios, the simulation time is set to 30 seconds with measurement frequency at 12.5Hz. To assess validity in challenging propagation environments, the \ac{los} measurements only exist in the first 6 seconds. We generate ground truth trajectories using the motion model in \eqref{eq:motionmodel}, with motion parameters $\sigma_x=\sigma_y=0.5\rm m/s^2$. Each trajectory undergoes 10 repetitions, during which distinct \ac{mpc} measurements are generated using the measurement model in Section \ref{sec:wirmeas} for each repetition. Measurement noise parameters are set to $\sigma_{\rm d}=0.3\rm m$,$ \sigma_{\rm \theta}=4\deg$ and $\sigma^{(0)}_d=0.05\rm m, \sigma^{(0)}_{\theta}=2\deg$, with $P_{{\rm D},k,j}^{(i)}=0.95$ within a fixed-range \ac{fov} for the agent and $\lambda_{\rm c}=0.02$. For the filter setup, we set the particle number to $N=1000$. The filter parameters associated with the measurement model and motion model, including the particles' initial states, align with those employed in the data generation step, and we set the noise parameter for the clock bias as $\sigma_{\rm b}=0.01\rm m/s$. 
For birth \acp{gc}' parameters we discussed in \eqref{eq:birthconstraint} and \eqref{eq:birthcov} in Section \ref{sec:phd_predict}, we set $\gamma=0.7, \zeta=0.1, \iota=0.5, \xi=0.3$ and $\alpha_{\rm B}=0.01$. Specifically, in the first scenario, utilizing 10 trajectories and 100 simulations, the agent initial position and velocity are $(0,0)\rm m$ and $(1,0)\rm m/s$ respectively with $b_0=0.3\rm m$. 
 In particular, the agent's \ac{fov} is 35m. In the second scenario, which encompasses 5 trajectories and 50 simulations, the initial state and velocity fo the agent are designated as $(8,-10)\rm m$ and $(0,1)\rm m/s$, respectively, with $b_0=0.3\rm m$,  and the \ac{fov} is set to $25\rm m$.

To evaluate the effectiveness of our method,  we compare it with the approach presented in \cite{KimGraGaoBatKimWym:J20}, where two different \ac{phd} filters are used for handing the reflection and scattering \acp{mpc} separately. We also compare our method with an \ac{ekf}-based method that only utilizes \ac{los} measurements as a performance baseline. The parameters in these methods align with those employed in the data generation step.
The estimated trajectories from one experiment are visualized in Fig. \ref{fig:simtraj1} and Fig. \ref{fig:simtraj2}. The \ac{cdf} of the localization error across all simulation runs is shown in Fig. \ref{fig:simcdf}, and the \acp{rmse} of the 2-D position estimates for all \acp{vt} are shown in Tables \ref{tab:vtrmse1} and \ref{tab:vtrmse2}. The time complexity of our method and the SLAM in \cite{KimGraGaoBatKimWym:J20} per iteration are both in the order of $\mathcal{O}(N_k\times |\mathcal{M}_k|\times |\mathcal{Z}_{k}^{{\rm NLOS}}|)$.  For scenarios 1 and 2, respectively, the average computation time per iteration for our method is 17.6s and 29.2s, compared to 32.2s and 48.9s for the method in \cite{KimGraGaoBatKimWym:J20}.\footnote{The simulations are conducted using MATLAB, running on an Intel Xeon E5-2697 v4 @ 2.30GHz, under the Linux CentOS Stream 8 operating system.}

Fig. \ref{fig:simresult} demonstrate the enhanced localization accuracy achieved by our method in both scenarios. 
{In the first scenario, our method achieves a localization accuracy with an \ac{rmse} of $0.25\rm m$, which is $74.1\%$ better than the \ac{slam} approach in \cite{KimGraGaoBatKimWym:J20} with an \ac{rmse}=$0.98\rm m$. It also outperforms the \ac{ekf} baseline, which has an \ac{rmse} of $7.80\rm m$, by $96.8\%$. Similarly, in  the second scenario, our method's \ac{rmse} of $0.18\rm m$ surpasses that of the  \ac{slam} in \cite{KimGraGaoBatKimWym:J20} (\ac{rmse} $0.83\rm m$) by $78.1\%$ and exceeds the \ac{ekf} baseline (\ac{rmse} $7.83\rm m$) by $97.7\%$.}
Compared with other methods, Fig. \ref{fig:simresult} also shows the robustness of our method in scenarios where \ac{los} is obscured, further demonstrating the robustness and  superiority of our method in complex multipath scenarios with a limited \ac{fov}.

In terms of environmental sensing, our method excels in achieving sub-meter estimation accuracy in both scenarios, as demonstrated in Table \ref{tab:vtrmse1} and \ref{tab:vtrmse2}. In contrast, the radio-\ac{slam} method in \cite{KimGraGaoBatKimWym:J20} cannot handle \acp{vt} influenced by both reflections and scatterings. As a result, it yields less accurate estimates for VT3,4 in scenario 1, and for VT6,7,8,9 in scenario 2.\footnote{
In Table \ref{tab:vtrmse1}, VT2 and VT4 share the same 2-D position, since the method in \cite{KimGraGaoBatKimWym:J20} considers only the 2-D positions of \acp{vt}, it achieves identical \acp{rmse} for two \acp{vt}.
In Table \ref{tab:vtrmse2}, VT8 and VT9 have identical \acp{rmse}, while VT3 exhibits a different \ac{rmse} for the method in \cite{KimGraGaoBatKimWym:J20}. The reasons are as follows.  Firstly, VT8 and VT9 share the same positions and propagation biases. Thus, in the \ac{GM}-\ac{phd} map, VT8 and VT9 are often represented by a single \ac{gc} with a weight close to 2. On the contrary, the \ac{gc} corresponding to VT3 has a weight close to 1, separating it from VT8 and VT9. }

{
\begin{remark}
The method in \cite{KimGraGaoBatKimWym:J20} employs a closed-form strategy for importance weighting, while our approach utilizes a multi-feature strategy, as shown in \eqref{eq:multifeature}. 
To conduct a brief ablation study, we test the same closed-form strategy from  \cite[Sec. 4.C]{KimGraGaoBatKimWym:J20} within our method. The resulting localization RMSEs for our method are $0.52\rm m$ and $0.48\rm m$ for scenarios 1 and 2, respectively. This represents a reduction in localization RMSE of 46.7\% and 42.2\% compared to the method in \cite{KimGraGaoBatKimWym:J20} for scenarios 1 and 2.
\end{remark}
}

\begin{table}[t]
    \centering
    
    \vspace{-1em}
    \caption{The \acp{rmse} of \ac{vt} estimates in scenario 1}
    \begin{tabular}{m{6em}m{4em}m{4em}m{4em}m{4em}} \toprule
    Method & VT1  & VT2  & VT3  & VT4  \\ \midrule
    Ours & 0.40 m& 0.39 m& 0.66 m & 0.46 m  \\ 
    SLAM in \cite{KimGraGaoBatKimWym:J20} & 1.77 m &	6.28 m &	5.11 m & 6.28 m \\ \bottomrule
    \end{tabular}
    \label{tab:vtrmse1}
    \vspace{-1em}
\end{table}

\section{Conclusions}\label{sec:conclude}
In this paper, we delivered high-accuracy localization and environmental sensing in an efficient and  robust manner. The key to our method is the unified characterization and utilization of different types of \acp{mpc} via an \ac{rfs}, which enables us to forge the requirement of identifying \ac{mpc} types prior to filtering and eliminates the necessity for employing multiple \ac{mpc} tracking filters. Moreover, our method also can handle \acp{mpc} influenced by both reflections and scatterings. The simulation results demonstrated that our method excels in  both robustness and effectiveness, particularly in complex environments where the \ac{los} path is obscured and in situations involving clutter and missed detections of \ac{mpc} measurements.

\begin{table}[t]
    \centering
    
    \vspace{-1em}
    \caption{The \acp{rmse} of \ac{vt} estimates in scenario 2}
    \begin{tabular}{m{6em}m{4em}m{4em}m{5em}m{5em}} \toprule
    Method & VT1  & VT2 & VT3  & VT4,5  \\ \midrule
    Ours & 0.82 m & 0.92 m & 0.15 m& 0.97 m \\ 
    SLAM in \cite{KimGraGaoBatKimWym:J20} & 1.68 m &	1.45 m &	0.98 m & 1.55 m \\ \toprule
    Method & VT6  & VT7  & VT8,9  &\\ \midrule
    Ours & 1.39 m & 0.29 m & 0.19 m &  \\ 
    SLAM in \cite{KimGraGaoBatKimWym:J20} & 4.54 m &	4.52 m &	5.23 m & \\\bottomrule
    \end{tabular}
    \label{tab:vtrmse2}
    \vspace{-1em}
\end{table}

\appendices


\bibliographystyle{IEEEtran}

\bibliography{IEEEAbrv,StringDefinitions,SGroupDefinition,SGroup}

\ifCLASSOPTIONcaptionsoff
  \newpage
\fi
 \newpage
 
\end{document}